\documentclass[aps,prl,twocolumn,showpacs,groupedaddress,reprint,amsmath,amssymb]{revtex4-1}

\usepackage{graphicx}
\usepackage[english]{babel}
\usepackage{array}
\usepackage{grffile} 

\usepackage{bm}
\renewcommand{\v}[1]{\bm{ #1 }}

\begin{document}

\newcolumntype{C}{>{$}c<{$}}
\newcolumntype{R}{>{$}r<{$}}
\newcolumntype{L}{>{$}l<{$}}


\title{Bayesian inference of the resonance content of $p(\gamma,K^{+})\Lambda$} 



\author{Lesley~De~Cruz}
\email[]{Lesley.DeCruz@UGent.be}
\affiliation{Department of Physics and Astronomy, Ghent University, Proeftuinstraat 86, B-9000 Gent, Belgium}

\author{Tom~Vrancx}
\affiliation{Department of Physics and Astronomy, Ghent University, Proeftuinstraat 86, B-9000 Gent, Belgium}

\author{Pieter~Vancraeyveld}
\affiliation{Department of Physics and Astronomy, Ghent University, Proeftuinstraat 86, B-9000 Gent, Belgium}

\author{Jan~Ryckebusch}
\email[]{Jan.Ryckebusch@UGent.be}
\affiliation{Department of Physics and Astronomy, Ghent University, Proeftuinstraat 86, B-9000 Gent, Belgium}


\date{\today}

\begin{abstract}
A Bayesian analysis of the world's $p(\gamma,K^+)\Lambda$ data is presented. From the proposed selection of 11 resonances, we find that the following nucleon resonances have the highest probability of contributing to the reaction: $S_{11}(1535)$, $S_{11}(1650)$, $F_{15}(1680)$, $P_{13}(1720)$, $D_{13}(1900)$, $P_{13}(1900)$, $P_{11}(1900)$, and $F_{15}(2000)$. We adopt a Regge-plus-resonance framework featuring consistent couplings for nucleon resonances up to spin $J =5/2$. We evaluate all possible combinations of 11 candidate resonances. The best model is selected from the 2048 model variants by calculating the Bayesian evidence values against the world's $p(\gamma,K^+)\Lambda$ data. 
\end{abstract}

\pacs{14.20.Gk, 14.40.Df, 11.55.Jy}

%
%
%

\maketitle 

A thorough knowledge of the nucleon-resonance ($N^\ast$) content of open-strangeness production reactions could dramatically improve our understanding of the nucleon's structure. Indeed, it provides a test-bed for the predicted $N^\ast$ spectra from competing baryon models \cite{loring-2001-nonstrange,ferretti-2011}. Despite being the subject of numerous analyses, the set of $N^{\ast}$s that contribute to $p (\gamma,K^+) \Lambda$ is not unambiguously determined. The Particle Data Group (PDG) \cite{nakamura-2010} lists four resonances with a fair evidence of existence in the $K^{+}\Lambda$ decay channel. Of these, only the $S_{11}(1650)$ has a three-star status which corresponds to a very likely contribution to this channel \cite{nakamura-2010}. This is reflected by the often contradictory outcomes of different analyses on which the PDG ratings are based. This disparity is illustrated in Table \ref{tab:othermodels}. The persistent lack of consensus, despite the increasing availability of $p(\gamma,K^{+})\Lambda$ data, can be attributed in part to the important role played by non-resonant dynamics.

\begin{table}[tbph]
\footnotesize
\caption{The sets of $N^{\ast}$s included in various pseudoscalar meson photoproduction analyses, compared to the results of this work. The nomenclature $L_{2I,2J}(M_{N^{\ast}})$ is used, where $L$ is the orbital angular momentum of the $\pi N$ partial wave, $I$ is the isospin, $J$ is the spin and $M_{N^{\ast}}$ is the mass of the resonance. The overall PDG ratings are given for each $N^{\ast}$. Missing states predicted by constituent quark models are denoted with $m$. Along with the $P_{11}(1440)$ and $D_{13}(1520)$, the $N^{\ast}$s with $J \geq 7/2$ are not considered in this work.
\label{tab:othermodels}
}
\begin{tabular}{|r|c|c|c|c|c|c|c|c|c|c|c|c|c|c|}
\hline Analysis & \rotatebox{90} {$P_{11}(1440)\; {\star}{\star}{\star}{\star}$} & \rotatebox{90} {$D_{13}(1520)\; {\star}{\star}{\star}{\star}$} & \rotatebox{90} {$S_{11}(1535)\; {\star}{\star}{\star}{\star}$} & \rotatebox{90} {$S_{11}(1650)\; {\star}{\star}{\star}{\star}$} & \rotatebox{90} {$D_{15}(1675)\; {\star}{\star}{\star}{\star}$} & \rotatebox{90} {$F_{15}(1680)\; {\star}{\star}{\star}{\star}$} & \rotatebox{90} {$D_{13}(1700)\; {\star}{\star}{\star}$} & \rotatebox{90} {$P_{11}(1710)\; {\star}{\star}{\star}$} & \rotatebox{90} {$P_{13}(1720)\; {\star}{\star}{\star}{\star}$} & \rotatebox{90} {$D_{13}(1900) \; m$} & \rotatebox{90} {$P_{13}(1900)\; {\star}{\star}$} & \rotatebox{90} {$P_{11}(1900)\; m$} & \rotatebox{90} {$F_{15}(2000)\; {\star}{\star}{\star}$} & \rotatebox{90} {$J \geq 7/2$}\\
\hline
RPR-2011 & & & \checkmark & \checkmark & & \checkmark & & & \checkmark & \checkmark & \checkmark & \checkmark & \checkmark &\\
B-G \cite{anisovich-2010} & \checkmark & \checkmark & \checkmark & \checkmark & \checkmark & \checkmark & & \checkmark & \checkmark & & & & & \checkmark\\
EBAC-DCC \cite{juliadiaz-2006} & & \checkmark & \checkmark & \checkmark & & & & & & \checkmark & \checkmark & & &\\
Gent-Isobar \cite{ireland-2004} & & & & \checkmark & & & & \checkmark & \checkmark & & & \checkmark & &\\
Giessen \cite{shklyar-2005} & & & & \checkmark & & & & \checkmark & \checkmark & & \checkmark & & &\\
KaonMAID \cite{mart-1999} & & & & \checkmark & & & & \checkmark & \checkmark & \checkmark & & & &\\
RPR-2007 \cite{corthals-2006} & & & & \checkmark & & & & \checkmark & \checkmark & \checkmark & \checkmark & & &\\
Saclay-Lyon \cite{david-1995} & \checkmark & & & & \checkmark & & & & \checkmark & & & & &\\
SAID \cite{arndt-2006} & \checkmark & \checkmark & \checkmark & \checkmark & \checkmark & \checkmark & & & \checkmark & & & & \checkmark & \checkmark\\
US/SSL \cite{usov-2006,shyam-2010} & \checkmark & \checkmark & \checkmark & \checkmark & & & \checkmark & \checkmark & \checkmark & & & & &\\
\hline
\end{tabular}

\end{table}
  
The criterion to determine whether a resonance contributes significantly or insignificantly varies among different analyses. In this Letter, we wish to address this issue in a statistically solid way, using Bayesian inference.
The past decade has seen the development of advanced coupled-channels (CC) models \cite{shklyar-2005,juliadiaz-2006,anisovich-2010,usov-2006,shyam-2010}. The effect of channel openings has been identified as playing an important role in the reaction dynamics \cite{suzuki-2009}. As Bayesian inference requires non-trivial numerical computations in the parameter space, to date it can only be done in an efficient single-channel reaction model, which does not capture the full complexity of CC models. We perform an analysis using a set of nucleon resonances that are likely to contribute to $p(\gamma,K^+)\Lambda$, within the Regge-plus-resonance (RPR) model \cite{corthals-2006,corthals-2007a,corthals-2007b}. The RPR model is devised as a unified description of both the high-energy region ($\sqrt{s} \gtrsim$ 2.5 GeV), where the differential cross section is forward peaked, and the resonance region ($\sqrt{s} \lesssim$ 2.5 GeV). In the RPR approach, the high-energy region is described using a Regge model. It is based on the exchange of the $K^{+}(494)$ and $K^{\ast+}(892)$ trajectories in the $t$-channel and is parametrized by three coupling strengths and two phases \cite{guidal-1997a}. In the resonance region, the Regge model provides a fair description of the elusive background. By coherently adding the $s$-channel nucleon-resonance contributions in this energy region, one obtains a description of the electromagnetic $K^{+}\Lambda$ production process for photon energies from threshold up to 16~GeV \cite{corthals-2006}.

Our formalism makes use of the recently suggested consistent couplings for resonances with $J=3/2$ and $J=5/2$ \cite{vrancx-2011}. This means that all spurious degrees of
freedom due to the lower-spin components are removed from the $J\geq3/2$ propagators. In addition, the couplings of all resonances with $J\geq3/2$ can be described by a mere two parameters. A spin-dependent multidipole-Gauss hadronic form factor (FF) \cite{vrancx-2011} is employed to regularize the resonance contributions beyond the $N^{\ast}$ pole ($\sqrt{s} > M_{N^\ast}$). In order to minimize the number of parameters, we adopt one
common cut-off value for the hadronic FF for all $N^{\ast}$s. 

The reggeized background is constrained using photoproduction data above the resonance region. In pion photoproduction \cite{sibirtsev-2007} the resonance region extends to $\sqrt{s} \approx$ 2.5~GeV. In previous work \cite{decruz-2010,boyarski-1969, quinn-1979, vogel-1972}, the 72 $p(\gamma,K^+)\Lambda$ data points with $\sqrt{s} >$ 3~GeV were employed to determine the parameters of the reggeized background. The CLAS collaboration has recently published $p(\gamma,K^+)\Lambda$ data for $\sqrt{s} > $ 2.5~GeV \cite{mccracken-2009}. The CLAS data are inconsistent with those collected in the sixties and seventies \cite{dey-2011}. Similar discrepancies were found in other pseudoscalar meson photoproduction channels \cite{dey-2011}. For the analysis presented here, we have used a subset of the recent CLAS data for which 2.6~GeV $< \sqrt{s} <$ 2.84~GeV and kaon center-of-momentum angle $\cos{\theta_K^{\ast}}>0.35$ to constrain the reggeized background model. By means of a Bayesian analysis, analogous to the procedure described in Ref. \cite{decruz-2010}, we have determined the optimal background model variant which is dubbed Regge-2011. This background model features rotating phases for both trajectories, and positive tensor and vector coupling strengths for the $K^{\ast+}$ trajectory. The RPR amplitude is constructed from this background model by adding a set of $s$-channel contributions.

The challenge at hand is to determine which set of resonances gives rise to the most probable RPR model variant $M$, given the world's $p(\gamma,K^{+})\Lambda$ data of the last decade, $\left\lbrace d_k \right\rbrace$. The data include 3455 differential cross sections, 2241 single, and 452 double polarization observables \cite{corthals-2006,mccracken-2009,bradford-2007,lleres-2008}. The set of resonances can be determined by evaluating the conditional probability $ P(M|\left\lbrace d_k \right\rbrace) = P(\left\lbrace d_k \right\rbrace|M) \, P(M) / P(\left\lbrace d_k \right\rbrace)$ \cite{ireland-2010,decruz-2010} 
for each model variant $M$. The factor $P(\left\lbrace d_k \right\rbrace|M)$ is known as the Bayesian evidence $\mathcal{Z}$. The probability ratio of two different models $M_A$ and $M_B$, given the data set $\left\lbrace d_k \right\rbrace$ can be expressed as
\begin{align}
\frac{P(M_A|\left\lbrace d_k \right\rbrace)}{P(M_B|\left\lbrace d_k \right\rbrace)} &= \frac{ P(\left\lbrace d_k \right\rbrace| M_A)}{ P(\left\lbrace d_k \right\rbrace| M_B)} \; \frac{P(M_A)}{P(M_B)} = \frac{ \mathcal{Z}_A}{\mathcal{Z}_B} \; \frac{P(M_A)}{P(M_B)}.\label{eqn:evidenceratio}
\end{align}
As we have no prior preference for any of the models, the factor $P(M_A)/P(M_B)$ is set to one, and the probability ratio of Eq.~(\ref{eqn:evidenceratio}) can be expressed in terms of the evidence ratio (or, Bayes factor) $\mathcal{Z}_A/\mathcal{Z}_B$. The evidence is calculated by marginalizing over the model's parameters $\v{\alpha_M}$ \cite{sivia-2006,decruz-2010},
\begin{align}		
\mathcal{Z}  &= \int P(\left\lbrace d_k \right\rbrace,\v{\alpha_M}|M)\, d\v{\alpha_M} = \int \mathcal{L} (\v{\alpha_M}) \, \pi(\v{\alpha_M})\,d\v{\alpha_M}. \label{eqn:evidence}
\end{align}
The prior distribution $\pi(\v{\alpha_M}) = P(\v{\alpha_M}|\,M)$ is chosen to be a uniform distribution \cite{ireland-2010} between $-100$ and $+100$ for the coupling strengths. This choice is motivated by naturalness arguments: coupling strengths of 100 give rise to a total cross section exceeding 25 $\mu$b, thereby overshooting the measured $p(\gamma,K^{+})\Lambda$ by several factors.

The likelihood function $\mathcal{L} (\v{\alpha_M}) \equiv P(\left\lbrace d_k \right\rbrace|\,\v{\alpha_M},M)$ is parametrized by a chi-square distribution. In evaluating $\mathcal{L}(\v{\alpha_M})$, the customary estimate of the total squared error  of a data point is the sum of the squared systematic and statistical errors: $\sigma_\text{tot}^2 = \sigma_\text{sys}^2 + \sigma_\text{stat}^2$. Due to the non-Gaussian and correlated nature of the systematic errors, the use of a chi-square distribution underestimates the real errors and the resulting evidences $\mathcal{Z}$ \cite{decruz-phd}. The total error is underestimated by $\sqrt{2}$ if the two errors are equal. Furthermore, there are generally at least two different sources of systematic errors, which are also added quadratically. A more conservative calculation, where the errors are added linearly, yields as a total error
\begin{align}
\sigma^{\prime}_\text{tot} &= \sigma_\text{stat} + \sigma^{\prime}_\text{sys} \approx  \sigma_\text{stat} + \sqrt{2}\sigma_\text{sys} \approx \frac{1+\sqrt{2}}{\sqrt{2}}\sigma_{\text{tot}}. \label{eqn:sigmatotprime}
\end{align}
The replacement $\sigma_\text{tot} \to \sigma^{\prime}_\text{tot}$ boils down to rescaling the errors in the chi-square distribution with $c = \frac{1+\sqrt{2}}{\sqrt{2}}$. The bulk of $\mathcal{Z}$ is determined by $\max\left\lbrace \mathcal{L}(\v{\alpha_M})\right\rbrace$, so one can correct for this underestimate by considering the scaling behaviour of the chi-square distribution at $\chi^2_\text{min}$ if the error is multiplied by $c$. The following relation holds,
\begin{align}
\ln \dfrac{\mathcal{L}(\v{\alpha}_M)}{\mathcal{L}^{\prime}_c (\v{\alpha}_M)} & = (k-2)\ln{c} - \chi^2_R(\v{\alpha}_M) \frac{k}{2}\;\frac{c^2-1}{c^2},\label{eqn:likelihood-scaling}
\end{align}
where $\mathcal{L}^{\prime}_c (\v{\alpha}_M)$ is the chi-square distribution for which the errors have been multiplied by $c$. $k$ is the number of degrees of freedom of the chi-square distribution, \emph{i.e.} the number of data points minus the number of free parameters. This results in the following correction for the computed evidence $\mathcal{Z}$ of a model,
\begin{align}
\ln\mathcal{Z}^{\prime} &\approx \ln \mathcal{Z} -  (k-2)\ln{c} + \chi^2_{R,\text{min}} \frac{k}{2}\;\frac{c^2-1}{c^2},\label{eqn:Z-scaling}
\end{align}
where $\chi^2_{R,\text{min}}$ is the model's minimum reduced $\chi^2$ value.

Jeffreys' scale \cite{jeffreys-1961} associates the logarithm of the evidence ratio of Eq.~(\ref{eqn:evidenceratio}) with a qualitative statement on the relative probabilities for two models. It states that a value of $\Delta \ln\mathcal{Z} = \ln \left( \mathcal{Z}_A / \mathcal{Z}_B \right) \gtrsim 1$ corresponds with significant evidence in favor of the more probable model, whereas a value smaller than 1 is barely worth mentioning. A value larger than 2.5 is strong to very strong, and a value of 5 and larger is decisive. This scale is employed to decide which resonance set describes the $p(\gamma,K^{+})\Lambda$ data best with the RPR model.

The numerical evaluation of Eq.~(\ref{eqn:evidence}) is very cumbersome, as the bulk of the likelihood $\mathcal{L}(\v{\alpha_M})$ is usually concentrated in a very small region of the multidimensional parameter space. Therefore, we adopt a numerical procedure which includes different steps. First, we employ a genetic algorithm to locate the global optimum in the likelihood hypersurface. Next, the covariance matrix about the optimum is determined using the {\sc minos} routine in {\sc root}'s {\sc minuit} package \cite{antcheva-2011}. Finally, the {\sc vegas} algorithm \cite{GSL} is adopted to calculate the evidence integral within the error boundaries determined by {\sc minos}. This localized integration yields a first estimate for the evidences $\mathcal{Z}$. As we are dealing with 6148 data points, the likelihood is apt to be unimodal and peaked in a small region in parameter space. Moreover, the chi-square distribution $\mathcal{L}(\v{\alpha_M})$ falls off very steeply with increasing $\chi^2$. Therefore, the volume about the global maximum in the likelihood surface effectively represents the bulk of the integral.

We have evaluated the evidence integral $\mathcal{Z}^{\prime}$ for all model variants corresponding to combinations of the nucleon resonances listed in Table \ref{tab:othermodels}. We consider established resonances, for which substantial experimental evidence exists: $S_{11}(1535)$, $S_{11}(1650)$, $D_{15}(1675)$, $F_{15}(1680)$, $D_{13}(1700)$, $P_{11}(1710)$, and $P_{13}(1720)$. The less-established $P_{13}(1900)$ \cite{shklyar-2005,juliadiaz-2006} and $F_{15}(2000)$ resonances are also included. The $P_{11}(1440)$ and $D_{13}(1520)$, which have masses significantly below threshold, are not considered in our single-channel formalism. We include the ``missing''  $D_{13}(1900)$ and $P_{11}(1900)$ resonances. Both have been identified by at least one analysis as contributing to the $K^{+}\Lambda$ channel \cite{mart-1999,shklyar-2005,juliadiaz-2006,ireland-2004}. In a single-channel reaction model it is customary to introduce $N^{\ast}$ propagators with a single pole in the complex plane. Thereby, the dynamical origin of the $N^{\ast}$ \cite{suzuki-2009} is approximated by an effective mass and width. We have adopted the PDG values for the Breit-Wigner masses and widths if available.

\begin{figure}[tpb]
 \centering
\includegraphics[width=\columnwidth]{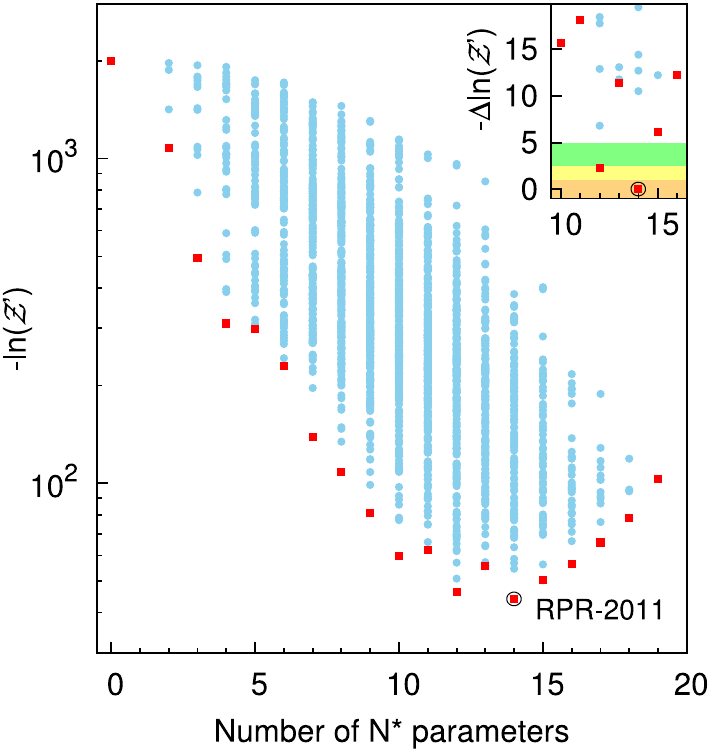}
  \caption{(color online). The evidence values of the 2048 model variants in the RPR model space (blue circles), as a function of the number of free $N^{\ast}$ parameters. The smaller the value of $-\ln\mathcal{Z}^{\prime}$ the higher the evidence. The best model for a fixed number of parameters is indicated with a red square, the overall best model (RPR-2011) with an open circle. Top right inset: evidence ratios relative to RPR-2011 $-\ln{(\mathcal{Z}^{\prime}_i / \mathcal{Z}^{\prime}_{\text{RPR-2011}})}$ for the models with the lowest $-\ln\mathcal{Z}^{\prime}$. The color coding refers to Jeffreys' scale: barely worth mentioning (orange), significant (yellow), strong to very strong (green) and decisive (white). 
 }
 \label{fig:bestdots} 
\end{figure}

We consider all possible combinations of the 11 proposed resonances, which yields 2048 model variants. In Fig.~\ref{fig:bestdots}, the computed evidence values $-\ln\mathcal{Z}^{\prime}$ are displayed in the evidence map of the RPR model space. The number of free parameters per resonance is 1 for spin-1/2 and 2 for higher-spin resonances. Points in the same column represent model variants that have the same number of free parameters, but have a different selection of resonances. This figure illustrates that increasing the number of $N^{\ast}$ parameters does not necessarily result in an improved evidence. The $\ln\mathcal{Z}^{\prime}$ does improve by almost two orders of magnitude by including $N^{\ast}$s. 

The model with the highest evidence value has 14 $N^{\ast}$ parameters and features the resonances $S_{11}(1535)$, $S_{11}(1650)$, $F_{15}(1680)$, $P_{13}(1720)$, $D_{13}(1900)$, $P_{13}(1900)$, $P_{11}(1900)$, and $F_{15}(2000)$. This model is dubbed RPR-2011. A comparison with the second-best model, which has 12 $N^{\ast}$ parameters and does not feature the missing $D_{13}(1900)$ resonance, yields a difference of $\Delta \ln \mathcal{Z} = 2.3$, corresponding with significant to strong evidence in favor of the RPR-2011 model. 

\begin{figure*}[tpb]
 \centering
\includegraphics[width=\textwidth]{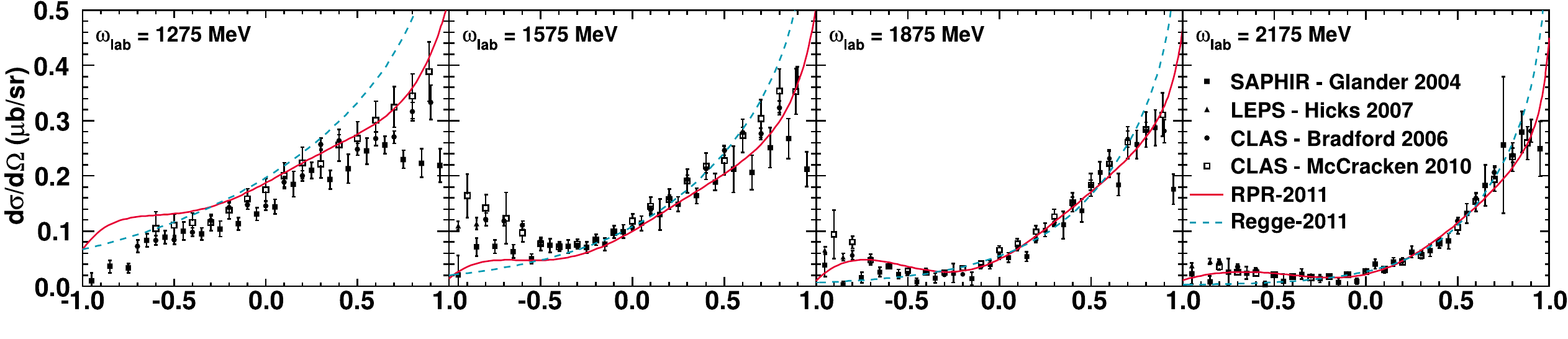}
\includegraphics[width=\textwidth]{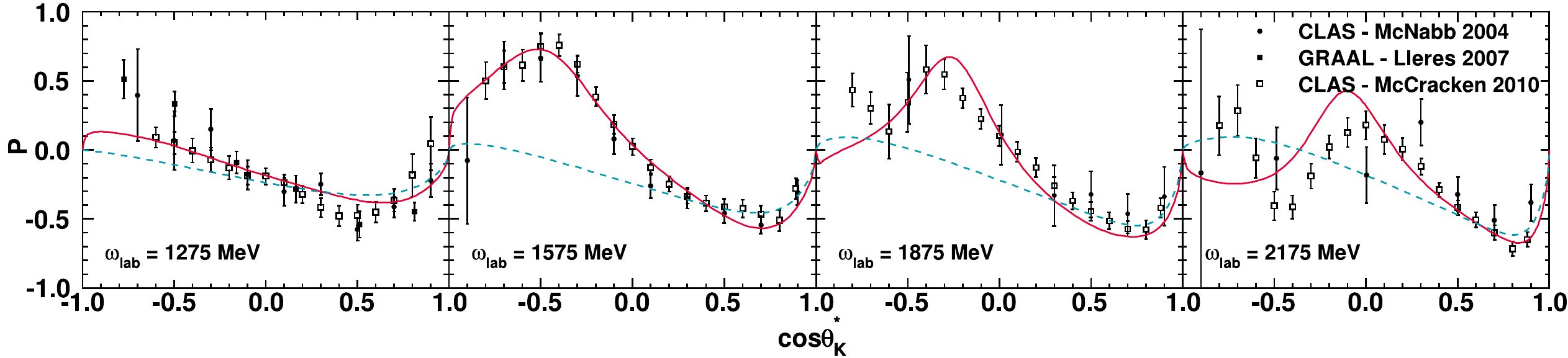}
 \caption{(color online). The angular dependence of the $p(\gamma,K^{+})\Lambda$ differential cross section and recoil polarization $P$ at various $\omega_{\text{lab}}$. The full red line represents the RPR-2011 model and the blue dashed line corresponds to the reggeized background model Regge-2011. Data are from Refs. \cite{bradford-2006,glander-2004,hicks-2007,mccracken-2009,mcnabb-2004,lleres-2007}.}
 \label{fig:dcs} 
\end{figure*}

A selection of the $p(\gamma,K^{+})\Lambda$ differential cross sections and recoil polarizations $P$ as calculated by the RPR-2011 model and the Regge-2011 background model are presented in Fig.~\ref{fig:dcs}. The figure illustrates that while the differential cross section is dominated by the background amplitude, polarization observables can be highly sensitive to $N^{\ast}$ contributions. The highest sensitivity to the $N^{\ast}$ contributions can be observed at backward kaon angles $\theta_K^{\ast}$.

We use the 2048 evidences of Fig.~\ref{fig:bestdots} to determine the conditional probability of the individual resonances, given the 6148 measured $p(\gamma,K^+)\Lambda$ observables,
\begin{align}
 P\left(R \; | \left\lbrace d_k \right\rbrace \right) 	& = \sum_{M_i | R \in M_i} P\left(\left\lbrace d_k \right\rbrace|M_i\right) \; \frac{P\left(M_i\right)}{P\left(\left\lbrace d_k \right\rbrace\right)}, \label{eqn:bayessumres}
 \end{align}
where the summation includes the $n$ model variants $M_i$ containing $R$. The second factor on the right hand side, $P\left(M_i\right) / P\left(\left\lbrace d_k \right\rbrace\right)$, is equal for all models $M_i$, so it drops out of the probability ratio $ P\left(R_1 \; | \left\lbrace d_k \right\rbrace \right) /  P\left(R_2 \; | \left\lbrace d_k \right\rbrace \right)$ of two resonances. 

\begin{figure}[tbp]
 \centering
 \includegraphics[width=\columnwidth]{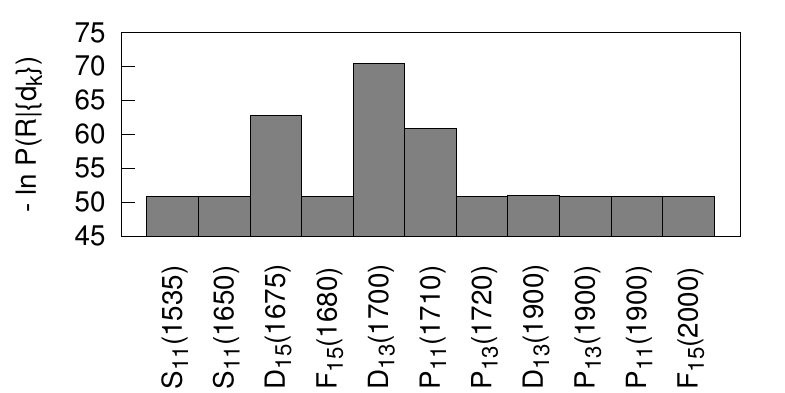}
  \caption{The relative resonance probabilities $- \ln P\left(R \; | \left\lbrace d_k \right\rbrace \right)$ of Eq.~(\ref{eqn:bayessumres}), computed from the evidences $\mathcal{Z}^{\prime}$ of Eq.~(\ref{eqn:Z-scaling}),  with $P\left(M_i\right) / P\left(\left\lbrace d_k \right\rbrace\right) = 1/n$. The three resonances that stand out are the ones that are not included in RPR-2011.}
 \label{fig:estimate}
\end{figure}

The results of Eq.~(\ref{eqn:bayessumres}) with $P\left(M_i\right) / P\left(\left\lbrace d_k \right\rbrace\right)$ set equal to 1/$n$ are shown in Fig.~\ref{fig:estimate}. This reveals that the resonances which have the highest probability of contributing to $p(\gamma,K^{+})\Lambda$ are those that constitute the resonance set of RPR-2011. This set features the three resonances with a mass around 1900~MeV that are predicted by constituent quark models \cite{loring-2001-nonstrange}, but not by quark-diquark models \cite{ferretti-2011}. Moreover, we find no significant contribution of the $P_{11}(1710)$ resonance to the $p(\gamma,K^{+})\Lambda$ reaction. In the latest SAID analyses \cite{arndt-2006,workman-2011}, this resonance was not needed for the description of $\pi N$ scattering either. The $P_{11}(1710)$'s negligible coupling to $\pi N$ and its absence in reactions where the $\pi \pi N$ channels are not relevant can be attributed to it being a resonance in the $\pi \pi N$ system \cite{khemchandani-2008}.

Summarizing, we have addressed the issue of investigating the resonance content of $p(\gamma,K^+)\Lambda$. This channel is known to receive large non-resonant contributions which complicates the extraction of $N^{\ast}$ information. The non-resonant dynamics can be effectively handled in a Regge formalism with a mere three parameters and two phases. The efficiency of the RPR model has enabled us to perform a Bayesian analysis. From a proposed set of 11 $N^{\ast}$s, we have identified the 8 $N^{\ast}$s with a high conditional probability of contributing to $p(\gamma,K^+)\Lambda$.

Bayesian inference has the power to reduce the bias in identifying the $N^{\ast}$ content in more advanced reaction models, but its dimensional curse leads to computational hurdles which cannot be overcome to date.

\begin{acknowledgments}
This research was financed by the Flemish Research Foundation (FWO Vlaanderen). The calculations were carried out on the Stevin Supercomputer Infrastructure at Ghent University, funded by Ghent University, the Hercules Foundation and the Flemish Government -- department EWI.
\end{acknowledgments}

\bibliography{klambda}

\begin{thebibliography}{10}%
\makeatletter
\providecommand \@ifxundefined [1]{%
 \ifx #1\undefined \expandafter \@firstoftwo
 \else \expandafter \@secondoftwo
\fi
}%
\providecommand \@ifnum [1]{%
 \ifnum #1\expandafter \@firstoftwo
 \else \expandafter \@secondoftwo
\fi
}%
\providecommand \enquote [1]{``#1''}%
\providecommand \bibnamefont  [1]{#1}%
\providecommand \bibfnamefont [1]{#1}%
\providecommand \citenamefont [1]{#1}%
\providecommand\href[0]{\@sanitize\@href}%
\providecommand\@href[1]{\endgroup\@@startlink{#1}\endgroup\@@href}%
\providecommand\@@href[1]{#1\@@endlink}%
\providecommand \@sanitize [0]{\begingroup\catcode`\&12\catcode`\#12\relax}%
\@ifxundefined \pdfoutput {\@firstoftwo}{%
 \@ifnum{\z@=\pdfoutput}{\@firstoftwo}{\@secondoftwo}%
}{%
 \providecommand\@@startlink[1]{\leavevmode\special{html:<a href="#1">}}%
 \providecommand\@@endlink[0]{\special{html:</a>}}%
}{%
 \providecommand\@@startlink[1]{%
  \leavevmode
  \pdfstartlink
   attr{/Border[0 0 1 ]/H/I/C[0 1 1]}%
   user{/Subtype/Link/A<</Type/Action/S/URI/URI(#1)>>}%
  \relax
 }%
 \providecommand\@@endlink[0]{\pdfendlink}%
}%
\providecommand \url  [0]{\begingroup\@sanitize \@url }%
\providecommand \@url [1]{\endgroup\@href {#1}{\urlprefix}}%
\providecommand \urlprefix [0]{URL }%
\providecommand \Eprint[0]{\href }%
\@ifxundefined \urlstyle {%
  \providecommand \doi [1]{doi:\discretionary{}{}{}#1}%
}{%
  \providecommand \doi [0]{doi:\discretionary{}{}{}\begingroup
  \urlstyle{rm}\Url }%
}%
\providecommand \doibase [0]{http://dx.doi.org/}%
\providecommand \Doi[1]{\href{\doibase#1}}%
\providecommand \bibAnnote [3]{%
  \BibitemShut{#1}%
  \begin{quotation}\noindent
    \textsc{Key:}\ #2\\\textsc{Annotation:}\ #3%
  \end{quotation}%
}%
\providecommand \bibAnnoteFile [2]{%
  \IfFileExists{#2}{\bibAnnote {#1} {#2} {\input{#2}}}{}%
}%
\providecommand \typeout [0]{\immediate \write \m@ne }%
\providecommand \selectlanguage [0]{\@gobble}%
\providecommand \bibinfo [0]{\@secondoftwo}%
\providecommand \bibfield [0]{\@secondoftwo}%
\providecommand \translation [1]{[#1]}%
\providecommand \BibitemOpen[0]{}%
\providecommand \bibitemStop [0]{}%
\providecommand \bibitemNoStop [0]{.\EOS\space}%
\providecommand \EOS [0]{\spacefactor3000\relax}%
\providecommand \BibitemShut [1]{\csname bibitem#1\endcsname}%
\bibitem{loring-2001-nonstrange}%
  \BibitemOpen
  \bibfield{author}{%
  \bibinfo {author} {\bibfnamefont{U.}~\bibnamefont{Loring}}, \bibinfo {author}
  {\bibfnamefont{B.~C.}\ \bibnamefont{Metsch}},\ and\ \bibinfo {author}
  {\bibfnamefont{H.~R.}\ \bibnamefont{Petry}},\ }%
  \bibfield{journal}{%
  \Doi{10.1007/s100500170105}{\bibinfo {journal} {Eur. Phys. J.}}\ }%
  \textbf{\bibinfo {volume} {A10}},\ \bibinfo {pages} {395} (\bibinfo {year}
  {2001}),\ \Eprint{http://arxiv.org/abs/hep-ph/0103289}{arXiv:hep-ph/0103289}%
  \bibAnnoteFile{NoStop}{loring-2001-nonstrange}%
\bibitem{ferretti-2011}%
  \BibitemOpen
  \bibfield{author}{%
  \bibinfo {author} {\bibfnamefont{J.}~\bibnamefont{Ferretti}}, \bibinfo
  {author} {\bibfnamefont{A.}~\bibnamefont{Vassallo}},\ and\ \bibinfo {author}
  {\bibfnamefont{E.}~\bibnamefont{Santopinto}},\ }%
  \bibfield{journal}{%
  \Doi{10.1103/PhysRevC.83.065204}{\bibinfo {journal} {Phys. Rev.}}\ }%
  \textbf{\bibinfo {volume} {C83}},\ \bibinfo {pages} {065204} (\bibinfo {year}
  {2011})%
  \bibAnnoteFile{NoStop}{ferretti-2011}%
\bibitem{nakamura-2010}%
  \BibitemOpen
  \bibfield{author}{%
  \bibinfo {author} {\bibfnamefont{K.}~\bibnamefont{Nakamura}} \emph{et~al.}
  (\bibinfo {collaboration} {Particle Data Group}),\ }%
  \bibfield{journal}{%
  \Doi{10.1088/0954-3899/37/7A/075021}{\bibinfo {journal} {J. Phys.}}\ }%
  \textbf{\bibinfo {volume} {G37}},\ \bibinfo {pages} {075021} (\bibinfo {year}
  {2010}),\ \url{http://pdg.lbl.gov}%
  \bibAnnoteFile{NoStop}{nakamura-2010}%
\bibitem{anisovich-2010}%
  \BibitemOpen
  \bibfield{author}{%
  \bibinfo {author} {\bibfnamefont{A.~V.}\ \bibnamefont{Anisovich}}, \bibinfo
  {author} {\bibfnamefont{E.}~\bibnamefont{Klempt}}, \bibinfo {author}
  {\bibfnamefont{V.~A.}\ \bibnamefont{Nikonov}}, \bibinfo {author}
  {\bibfnamefont{M.~A.}\ \bibnamefont{Matveev}}, \bibinfo {author}
  {\bibfnamefont{A.~V.}\ \bibnamefont{Sarantsev}}, \emph{et~al.},\ }%
  \bibfield{journal}{%
  \Doi{10.1140/epja/i2010-10950-x}{\bibinfo {journal} {Eur. Phys. J.}}\ }%
  \textbf{\bibinfo {volume} {A44}},\ \bibinfo {pages} {203} (\bibinfo {year}
  {2010}),\ \Eprint{http://arxiv.org/abs/0911.5277}{arXiv:0911.5277 [hep-ph]}%
  \bibAnnoteFile{NoStop}{anisovich-2010}%
\bibitem{juliadiaz-2006}%
  \BibitemOpen
  \bibfield{author}{%
  \bibinfo {author} {\bibfnamefont{B.}~\bibnamefont{Julia-Diaz}}, \bibinfo
  {author} {\bibfnamefont{B.}~\bibnamefont{Saghai}}, \bibinfo {author}
  {\bibfnamefont{T.~S.~H.}\ \bibnamefont{Lee}},\ and\ \bibinfo {author}
  {\bibfnamefont{F.}~\bibnamefont{Tabakin}},\ }%
  \bibfield{journal}{%
  \Doi{10.1103/PhysRevC.73.055204}{\bibinfo {journal} {Phys. Rev.}}\ }%
  \textbf{\bibinfo {volume} {C73}},\ \bibinfo {pages} {055204} (\bibinfo {year}
  {2006}),\
  \Eprint{http://arxiv.org/abs/nucl-th/0601053}{arXiv:nucl-th/0601053}%
  \bibAnnoteFile{NoStop}{juliadiaz-2006}%
\bibitem{ireland-2004}%
  \BibitemOpen
  \bibfield{author}{%
  \bibinfo {author} {\bibfnamefont{D.~G.}\ \bibnamefont{Ireland}}, \bibinfo
  {author} {\bibfnamefont{S.}~\bibnamefont{Janssen}},\ and\ \bibinfo {author}
  {\bibfnamefont{J.}~\bibnamefont{Ryckebusch}},\ }%
  \bibfield{journal}{%
  \Doi{10.1016/j.nuclphysa.2004.05.007}{\bibinfo {journal} {Nucl. Phys.}}\ }%
  \textbf{\bibinfo {volume} {A740}},\ \bibinfo {pages} {147} (\bibinfo {year}
  {2004})%
  \bibAnnoteFile{NoStop}{ireland-2004}%
\bibitem{shklyar-2005}%
  \BibitemOpen
  \bibfield{author}{%
  \bibinfo {author} {\bibfnamefont{V.}~\bibnamefont{Shklyar}}, \bibinfo
  {author} {\bibfnamefont{H.}~\bibnamefont{Lenske}},\ and\ \bibinfo {author}
  {\bibfnamefont{U.}~\bibnamefont{Mosel}},\ }%
  \bibfield{journal}{%
  \Doi{10.1103/PhysRevC.72.015210}{\bibinfo {journal} {Phys. Rev.}}\ }%
  \textbf{\bibinfo {volume} {C72}},\ \bibinfo {pages} {015210} (\bibinfo {year}
  {2005}),\
  \Eprint{http://arxiv.org/abs/nucl-th/0505010}{arXiv:nucl-th/0505010}%
  \bibAnnoteFile{NoStop}{shklyar-2005}%
\bibitem{mart-1999}%
  \BibitemOpen
  \bibfield{author}{%
  \bibinfo {author} {\bibfnamefont{T.}~\bibnamefont{Mart}}\ and\ \bibinfo
  {author} {\bibfnamefont{C.}~\bibnamefont{Bennhold}},\ }%
  \bibfield{journal}{%
  \Doi{10.1103/PhysRevC.61.012201}{\bibinfo {journal} {Phys. Rev.}}\ }%
  \textbf{\bibinfo {volume} {C61}},\ \bibinfo {pages} {012201} (\bibinfo {year}
  {1999}),\
  \Eprint{http://arxiv.org/abs/nucl-th/9906096}{arXiv:nucl-th/9906096}%
  \bibAnnoteFile{NoStop}{mart-1999}%
\bibitem{corthals-2006}%
  \BibitemOpen
  \bibfield{author}{%
  \bibinfo {author} {\bibfnamefont{T.}~\bibnamefont{Corthals}}, \bibinfo
  {author} {\bibfnamefont{J.}~\bibnamefont{Ryckebusch}},\ and\ \bibinfo
  {author} {\bibfnamefont{T.}~\bibnamefont{Van~Cauteren}},\ }%
  \bibfield{journal}{%
  \Doi{10.1103/PhysRevC.73.045207}{\bibinfo {journal} {Phys. Rev.}}\ }%
  \textbf{\bibinfo {volume} {C73}},\ \bibinfo {pages} {045207} (\bibinfo {year}
  {2006}),\
  \Eprint{http://arxiv.org/abs/nucl-th/0510056}{arXiv:nucl-th/0510056}%
  \bibAnnoteFile{NoStop}{corthals-2006}%
\bibitem{david-1995}%
  \BibitemOpen
  \bibfield{author}{%
  \bibinfo {author} {\bibfnamefont{J.~C.}\ \bibnamefont{David}}, \bibinfo
  {author} {\bibfnamefont{C.}~\bibnamefont{Fayard}}, \bibinfo {author}
  {\bibfnamefont{G.~H.}\ \bibnamefont{Lamot}},\ and\ \bibinfo {author}
  {\bibfnamefont{B.}~\bibnamefont{Saghai}},\ }%
  \bibfield{journal}{%
  \Doi{10.1103/PhysRevC.53.2613}{\bibinfo {journal} {Phys. Rev.}}\ }%
  \textbf{\bibinfo {volume} {C53}},\ \bibinfo {pages} {2613} (\bibinfo {year}
  {1996})%
  \bibAnnoteFile{NoStop}{david-1995}%
\bibitem{arndt-2006}%
  \BibitemOpen
  \bibfield{author}{%
  \bibinfo {author} {\bibfnamefont{R.~A.}\ \bibnamefont{Arndt}}, \bibinfo
  {author} {\bibfnamefont{W.~J.}\ \bibnamefont{Briscoe}}, \bibinfo {author}
  {\bibfnamefont{I.~I.}\ \bibnamefont{Strakovsky}},\ and\ \bibinfo {author}
  {\bibfnamefont{R.~L.}\ \bibnamefont{Workman}},\ }%
  \bibfield{journal}{%
  \Doi{10.1103/PhysRevC.74.045205}{\bibinfo {journal} {Phys. Rev.}}\ }%
  \textbf{\bibinfo {volume} {C74}},\ \bibinfo {pages} {045205} (\bibinfo {year}
  {2006}),\
  \Eprint{http://arxiv.org/abs/nucl-th/0605082}{arXiv:nucl-th/0605082}%
  \bibAnnoteFile{NoStop}{arndt-2006}%
\bibitem{usov-2006}%
  \BibitemOpen
  \bibfield{author}{%
  \bibinfo {author} {\bibfnamefont{A.}~\bibnamefont{Usov}}\ and\ \bibinfo
  {author} {\bibfnamefont{O.}~\bibnamefont{Scholten}},\ }%
  \bibfield{journal}{%
  \Doi{10.1103/PhysRevC.74.015205}{\bibinfo {journal} {Phys. Rev.}}\ }%
  \textbf{\bibinfo {volume} {C74}},\ \bibinfo {pages} {015205} (\bibinfo {year}
  {2006}),\
  \Eprint{http://arxiv.org/abs/nucl-th/0604009}{arXiv:nucl-th/0604009}%
  \bibAnnoteFile{NoStop}{usov-2006}%
\bibitem{shyam-2010}%
  \BibitemOpen
  \bibfield{author}{%
  \bibinfo {author} {\bibfnamefont{R.}~\bibnamefont{Shyam}}, \bibinfo {author}
  {\bibfnamefont{O.}~\bibnamefont{Scholten}},\ and\ \bibinfo {author}
  {\bibfnamefont{H.}~\bibnamefont{Lenske}},\ }%
  \bibfield{journal}{%
  \Doi{10.1103/PhysRevC.81.015204}{\bibinfo {journal} {Phys. Rev.}}\ }%
  \textbf{\bibinfo {volume} {C81}},\ \bibinfo {pages} {015204} (\bibinfo {year}
  {2010}),\ \Eprint{http://arxiv.org/abs/0911.3351}{arXiv:0911.3351 [hep-ph]}%
  \bibAnnoteFile{NoStop}{shyam-2010}%
\bibitem{suzuki-2009}%
  \BibitemOpen
  \bibfield{author}{%
  \bibinfo {author} {\bibfnamefont{N.}~\bibnamefont{Suzuki}}, \bibinfo {author}
  {\bibfnamefont{B.}~\bibnamefont{Julia-Diaz}}, \bibinfo {author}
  {\bibfnamefont{H.}~\bibnamefont{Kamano}}, \bibinfo {author}
  {\bibfnamefont{T.~S.~H.}\ \bibnamefont{Lee}}, \bibinfo {author}
  {\bibfnamefont{A.}~\bibnamefont{Matsuyama}}, \emph{et~al.},\ }%
  \bibfield{journal}{%
  \Doi{10.1103/PhysRevLett.104.042302}{\bibinfo {journal} {Phys.Rev.Lett.}}\ }%
  \textbf{\bibinfo {volume} {104}},\ \bibinfo {pages} {042302} (\bibinfo {year}
  {2010}),\ \Eprint{http://arxiv.org/abs/0909.1356}{arXiv:0909.1356 [nucl-th]}%
  \bibAnnoteFile{NoStop}{suzuki-2009}%
\bibitem{corthals-2007a}%
  \BibitemOpen
  \bibfield{author}{%
  \bibinfo {author} {\bibfnamefont{T.}~\bibnamefont{Corthals}}, \bibinfo
  {author} {\bibfnamefont{D.~G.}\ \bibnamefont{Ireland}}, \bibinfo {author}
  {\bibfnamefont{T.}~\bibnamefont{Van~Cauteren}},\ and\ \bibinfo {author}
  {\bibfnamefont{J.}~\bibnamefont{Ryckebusch}},\ }%
  \bibfield{journal}{%
  \Doi{10.1103/PhysRevC.75.045204}{\bibinfo {journal} {Phys. Rev.}}\ }%
  \textbf{\bibinfo {volume} {C75}},\ \bibinfo {pages} {045204} (\bibinfo {year}
  {2007}),\
  \Eprint{http://arxiv.org/abs/nucl-th/0612085}{arXiv:nucl-th/0612085}%
  \bibAnnoteFile{NoStop}{corthals-2007a}%
\bibitem{corthals-2007b}%
  \BibitemOpen
  \bibfield{author}{%
  \bibinfo {author} {\bibfnamefont{T.}~\bibnamefont{Corthals}}, \bibinfo
  {author} {\bibfnamefont{T.}~\bibnamefont{Van~Cauteren}}, \bibinfo {author}
  {\bibfnamefont{P.}~\bibnamefont{Vancraeyveld}}, \bibinfo {author}
  {\bibfnamefont{J.}~\bibnamefont{Ryckebusch}},\ and\ \bibinfo {author}
  {\bibfnamefont{D.~G.}\ \bibnamefont{Ireland}},\ }%
  \bibfield{journal}{%
  \bibinfo {journal} {Phys. Lett.}\ }%
  \textbf{\bibinfo {volume} {B656}},\ \bibinfo {pages} {186} (\bibinfo {year}
  {2007}),\ \Eprint{http://arxiv.org/abs/0704.3691}{arXiv:0704.3691 [nucl-th]}%
  \bibAnnoteFile{NoStop}{corthals-2007b}%
\bibitem{guidal-1997a}%
  \BibitemOpen
  \bibfield{author}{%
  \bibinfo {author} {\bibfnamefont{M.}~\bibnamefont{Guidal}}, \bibinfo {author}
  {\bibfnamefont{J.}~\bibnamefont{Laget}},\ and\ \bibinfo {author}
  {\bibfnamefont{M.}~\bibnamefont{Vanderhaeghen}},\ }%
  \bibfield{journal}{%
  \Doi{10.1016/S0375-9474(97)00612-X}{\bibinfo {journal} {Nucl. Phys.}}\ }%
  \textbf{\bibinfo {volume} {A627}},\ \bibinfo {pages} {645} (\bibinfo {year}
  {1997})%
  \bibAnnoteFile{NoStop}{guidal-1997a}%
\bibitem{vrancx-2011}%
  \BibitemOpen
  \bibfield{author}{%
  \bibinfo {author} {\bibfnamefont{T.}~\bibnamefont{Vrancx}}, \bibinfo {author}
  {\bibfnamefont{L.}~\bibnamefont{De~Cruz}}, \bibinfo {author}
  {\bibfnamefont{J.}~\bibnamefont{Ryckebusch}},\ and\ \bibinfo {author}
  {\bibfnamefont{P.}~\bibnamefont{Vancraeyveld}},\ }%
  \bibfield{journal}{%
  \Doi{10.1103/PhysRevC.84.045201}{\bibinfo {journal} {Phys. Rev.}}\ }%
  \textbf{\bibinfo {volume} {C84}},\ \bibinfo {pages} {045201} (\bibinfo {year}
  {2011}),\ \Eprint{http://arxiv.org/abs/1105.2688}{arXiv:1105.2688 [nucl-th]}%
  \bibAnnoteFile{NoStop}{vrancx-2011}%
\bibitem{sibirtsev-2007}%
  \BibitemOpen
  \bibfield{author}{%
  \bibinfo {author} {\bibfnamefont{A.}~\bibnamefont{Sibirtsev}}, \bibinfo
  {author} {\bibfnamefont{J.}~\bibnamefont{Haidenbauer}}, \bibinfo {author}
  {\bibfnamefont{S.}~\bibnamefont{Krewald}}, \bibinfo {author}
  {\bibfnamefont{T.~S.~H.}\ \bibnamefont{Lee}}, \bibinfo {author}
  {\bibfnamefont{U.-G.}\ \bibnamefont{Meissner}}, \emph{et~al.},\ }%
  \bibfield{journal}{%
  \Doi{10.1140/epja/i2007-10482-6}{\bibinfo {journal} {Eur. Phys. J.}}\ }%
  \textbf{\bibinfo {volume} {A34}},\ \bibinfo {pages} {49} (\bibinfo {year}
  {2007}),\ \Eprint{http://arxiv.org/abs/0706.0183}{arXiv:0706.0183 [nucl-th]}%
  \bibAnnoteFile{NoStop}{sibirtsev-2007}%
\bibitem{decruz-2010}%
  \BibitemOpen
  \bibfield{author}{%
  \bibinfo {author} {\bibfnamefont{L.}~\bibnamefont{De~Cruz}}, \bibinfo
  {author} {\bibfnamefont{D.~G.}\ \bibnamefont{Ireland}}, \bibinfo {author}
  {\bibfnamefont{P.}~\bibnamefont{Vancraeyveld}},\ and\ \bibinfo {author}
  {\bibfnamefont{J.}~\bibnamefont{Ryckebusch}},\ }%
  \bibfield{journal}{%
  \Doi{10.1016/j.physletb.2010.09.026}{\bibinfo {journal} {Phys. Lett.}}\ }%
  \textbf{\bibinfo {volume} {B694}},\ \bibinfo {pages} {33} (\bibinfo {year}
  {2010}),\ \Eprint{http://arxiv.org/abs/1004.0353}{arXiv:1004.0353 [nucl-th]}%
  \bibAnnoteFile{NoStop}{decruz-2010}%
\bibitem{boyarski-1969}%
  \BibitemOpen
  \bibfield{author}{%
  \bibinfo {author} {\bibfnamefont{A.}~\bibnamefont{Boyarski}} \emph{et~al.},\
  }%
  \bibfield{journal}{%
  \Doi{10.1103/PhysRevLett.22.1131}{\bibinfo {journal} {Phys. Rev. Lett.}}\ }%
  \textbf{\bibinfo {volume} {22}},\ \bibinfo {pages} {1131} (\bibinfo {year}
  {1969})%
  \bibAnnoteFile{NoStop}{boyarski-1969}%
\bibitem{quinn-1979}%
  \BibitemOpen
  \bibfield{author}{%
  \bibinfo {author} {\bibfnamefont{D.~J.}\ \bibnamefont{Quinn}} \emph{et~al.},\
  }%
  \bibfield{journal}{%
  \Doi{10.1103/PhysRevD.20.1553}{\bibinfo {journal} {Phys. Rev.}}\ }%
  \textbf{\bibinfo {volume} {D20}},\ \bibinfo {pages} {1553} (\bibinfo {year}
  {1979})%
  \bibAnnoteFile{NoStop}{quinn-1979}%
\bibitem{vogel-1972}%
  \BibitemOpen
  \bibfield{author}{%
  \bibinfo {author} {\bibfnamefont{G.}~\bibnamefont{Vogel}} \emph{et~al.},\ }%
  \bibfield{journal}{%
  \Doi{10.1016/0370-2693(72)90566-7}{\bibinfo {journal} {Phys. Lett.}}\ }%
  \textbf{\bibinfo {volume} {B40}},\ \bibinfo {pages} {513} (\bibinfo {year}
  {1972})%
  \bibAnnoteFile{NoStop}{vogel-1972}%
\bibitem{mccracken-2009}%
  \BibitemOpen
  \bibfield{author}{%
  \bibinfo {author} {\bibfnamefont{M.~E.}\ \bibnamefont{McCracken}}
  \emph{et~al.} (\bibinfo {collaboration} {CLAS}),\ }%
  \bibfield{journal}{%
  \Doi{10.1103/PhysRevC.81.025201}{\bibinfo {journal} {Phys. Rev.}}\ }%
  \textbf{\bibinfo {volume} {C81}},\ \bibinfo {pages} {025201} (\bibinfo {year}
  {2010}),\ \Eprint{http://arxiv.org/abs/0912.4274}{arXiv:0912.4274 [nucl-ex]}%
  \bibAnnoteFile{NoStop}{mccracken-2009}%
\bibitem{dey-2011}%
  \BibitemOpen
  \bibfield{author}{%
  \bibinfo {author} {\bibfnamefont{B.}~\bibnamefont{Dey}}\ and\ \bibinfo
  {author} {\bibfnamefont{C.~A.}\ \bibnamefont{Meyer}}}%
   (\bibinfo {year} {2011}),\
  \Eprint{http://arxiv.org/abs/1106.0479}{arXiv:1106.0479 [hep-ph]}%
  \bibAnnoteFile{NoStop}{dey-2011}%
\bibitem{bradford-2007}%
  \BibitemOpen
  \bibfield{author}{%
  \bibinfo {author} {\bibfnamefont{R.}~\bibnamefont{Bradford}} \emph{et~al.}
  (\bibinfo {collaboration} {CLAS}),\ }%
  \bibfield{journal}{%
  \Doi{10.1103/PhysRevC.75.035205}{\bibinfo {journal} {Phys. Rev.}}\ }%
  \textbf{\bibinfo {volume} {C75}},\ \bibinfo {pages} {035205} (\bibinfo {year}
  {2007}),\
  \Eprint{http://arxiv.org/abs/nucl-ex/0611034}{arXiv:nucl-ex/0611034}%
  \bibAnnoteFile{NoStop}{bradford-2007}%
\bibitem{lleres-2008}%
  \BibitemOpen
  \bibfield{author}{%
  \bibinfo {author} {\bibfnamefont{A.}~\bibnamefont{Lleres}} \emph{et~al.}
  (\bibinfo {collaboration} {GRAAL}),\ }%
  \bibfield{journal}{%
  \Doi{10.1140/epja/i2008-10713-4}{\bibinfo {journal} {Eur. Phys. J.}}\ }%
  \textbf{\bibinfo {volume} {A39}},\ \bibinfo {pages} {149} (\bibinfo {year}
  {2009}),\ \Eprint{http://arxiv.org/abs/0807.3839}{arXiv:0807.3839 [nucl-ex]}%
  \bibAnnoteFile{NoStop}{lleres-2008}%
\bibitem{ireland-2010}%
  \BibitemOpen
  \bibfield{author}{%
  \bibinfo {author} {\bibfnamefont{D.~G.}\ \bibnamefont{Ireland}},\ }%
  \bibfield{journal}{%
  \Doi{10.1103/PhysRevC.82.025204}{\bibinfo {journal} {Phys. Rev.}}\ }%
  \textbf{\bibinfo {volume} {C82}},\ \bibinfo {pages} {025204} (\bibinfo {year}
  {2010}),\ \Eprint{http://arxiv.org/abs/1004.5250}{arXiv:1004.5250 [hep-ph]}%
  \bibAnnoteFile{NoStop}{ireland-2010}%
\bibitem{sivia-2006}%
  \BibitemOpen
  \bibfield{author}{%
  \bibinfo {author} {\bibfnamefont{D.~S.}\ \bibnamefont{Sivia}}\ and\ \bibinfo
  {author} {\bibfnamefont{J.}~\bibnamefont{Skilling}},\ }%
  \emph{\bibinfo {title} {Data Analysis -- A Bayesian Tutorial}}\ (\bibinfo
  {publisher} {Oxford Science Publications},\ \bibinfo {year} {2006})%
  \bibAnnoteFile{NoStop}{sivia-2006}%
\bibitem{decruz-phd}%
  \BibitemOpen
  \bibfield{author}{%
  \bibinfo {author} {\bibfnamefont{L.}~\bibnamefont{De~Cruz}},\ }%
  Ph.D. thesis,\ \bibinfo {school} {Ghent University} (\bibinfo {year}
  {2011}),\ \url{http://inwpent5.ugent.be/papers/phdlesley.pdf}%
  \bibAnnoteFile{NoStop}{decruz-phd}%
\bibitem{jeffreys-1961}%
  \BibitemOpen
  \bibfield{author}{%
  \bibinfo {author} {\bibfnamefont{S.~H.}\ \bibnamefont{Jeffreys}},\ }%
  \emph{\bibinfo {title} {Theory of Probability}}\ (\bibinfo {publisher}
  {Oxford University Press},\ \bibinfo {year} {1961})%
  \bibAnnoteFile{NoStop}{jeffreys-1961}%
\bibitem{antcheva-2011}%
  \BibitemOpen
  \bibfield{author}{%
  \bibinfo {author} {\bibfnamefont{I.}~\bibnamefont{Antcheva}} \emph{et~al.},\
  }%
  \bibfield{journal}{%
  \Doi{10.1016/j.cpc.2011.02.008}{\bibinfo {journal} {Comput. Phys. Commun.}}\
  }%
  \textbf{\bibinfo {volume} {182}},\ \bibinfo {pages} {1384} (\bibinfo {year}
  {2011})%
  \bibAnnoteFile{NoStop}{antcheva-2011}%
\bibitem{GSL}%
  \BibitemOpen
  \bibfield{author}{%
  \bibinfo {author} {\bibfnamefont{M.}~\bibnamefont{Galassi}} \emph{et~al.},\
  }%
  \emph{\bibinfo {title} {{GNU Scientific Library Reference Manual - 3rd
  Edition}}}\ (\bibinfo {publisher} {Network Theory Ltd.},\ \bibinfo {year}
  {2009})\ \url{http://www.gnu.org/s/gsl/}%
  \bibAnnoteFile{NoStop}{GSL}%
\bibitem{bradford-2006}%
  \BibitemOpen
  \bibfield{author}{%
  \bibinfo {author} {\bibfnamefont{R.}~\bibnamefont{Bradford}} \emph{et~al.}
  (\bibinfo {collaboration} {CLAS}),\ }%
  \bibfield{journal}{%
  \Doi{10.1103/PhysRevC.73.035202}{\bibinfo {journal} {Phys. Rev.}}\ }%
  \textbf{\bibinfo {volume} {C73}},\ \bibinfo {pages} {035202} (\bibinfo {year}
  {2006}),\
  \Eprint{http://arxiv.org/abs/nucl-ex/0509033}{arXiv:nucl-ex/0509033}%
  \bibAnnoteFile{NoStop}{bradford-2006}%
\bibitem{glander-2004}%
  \BibitemOpen
  \bibfield{author}{%
  \bibinfo {author} {\bibfnamefont{K.~H.}\ \bibnamefont{Glander}}, \bibinfo
  {author} {\bibfnamefont{J.}~\bibnamefont{Barth}}, \bibinfo {author}
  {\bibfnamefont{W.}~\bibnamefont{Braun}}, \bibinfo {author}
  {\bibfnamefont{J.}~\bibnamefont{Hannappel}}, \bibinfo {author}
  {\bibfnamefont{N.}~\bibnamefont{Jopen}}, \emph{et~al.} (\bibinfo
  {collaboration} {SAPHIR}),\ }%
  \bibfield{journal}{%
  \Doi{10.1140/epja/i2003-10119-x}{\bibinfo {journal} {Eur. Phys.~J.}}\ }%
  \textbf{\bibinfo {volume} {A19}},\ \bibinfo {pages} {251} (\bibinfo {year}
  {2004}),\
  \Eprint{http://arxiv.org/abs/nucl-ex/0308025}{arXiv:nucl-ex/0308025}%
  \bibAnnoteFile{NoStop}{glander-2004}%
\bibitem{hicks-2007}%
  \BibitemOpen
  \bibfield{author}{%
  \bibinfo {author} {\bibfnamefont{K.}~\bibnamefont{Hicks}} \emph{et~al.}
  (\bibinfo {collaboration} {LEPS}),\ }%
  \bibfield{journal}{%
  \Doi{10.1103/PhysRevC.76.042201}{\bibinfo {journal} {Phys. Rev.}}\ }%
  \textbf{\bibinfo {volume} {C76}},\ \bibinfo {pages} {042201} (\bibinfo {year}
  {2007})%
  \bibAnnoteFile{NoStop}{hicks-2007}%
\bibitem{mcnabb-2004}%
  \BibitemOpen
  \bibfield{author}{%
  \bibinfo {author} {\bibfnamefont{J.~W.~C.}\ \bibnamefont{McNabb}}
  \emph{et~al.} (\bibinfo {collaboration} {CLAS}),\ }%
  \bibfield{journal}{%
  \Doi{10.1103/PhysRevC.69.042201}{\bibinfo {journal} {Phys. Rev.}}\ }%
  \textbf{\bibinfo {volume} {C69}},\ \bibinfo {pages} {042201} (\bibinfo {year}
  {2004}),\
  \Eprint{http://arxiv.org/abs/nucl-ex/0305028}{arXiv:nucl-ex/0305028}%
  \bibAnnoteFile{NoStop}{mcnabb-2004}%
\bibitem{lleres-2007}%
  \BibitemOpen
  \bibfield{author}{%
  \bibinfo {author} {\bibfnamefont{A.}~\bibnamefont{Lleres}}, \bibinfo {author}
  {\bibfnamefont{O.}~\bibnamefont{Bartalini}}, \bibinfo {author}
  {\bibfnamefont{V.}~\bibnamefont{Bellini}}, \bibinfo {author}
  {\bibfnamefont{J.~P.}\ \bibnamefont{Bocquet}}, \bibinfo {author}
  {\bibfnamefont{P.}~\bibnamefont{Calvat}}, \emph{et~al.} (\bibinfo
  {collaboration} {GRAAL}),\ }%
  \bibfield{journal}{%
  \Doi{10.1140/epja/i2006-10167-8}{\bibinfo {journal} {Eur. Phys. J.}}\ }%
  \textbf{\bibinfo {volume} {A31}},\ \bibinfo {pages} {79} (\bibinfo {year}
  {2007})%
  \bibAnnoteFile{NoStop}{lleres-2007}%
\bibitem{workman-2011}%
  \BibitemOpen
  \bibfield{author}{%
  \bibinfo {author} {\bibfnamefont{R.~L.}\ \bibnamefont{Workman}}, \bibinfo
  {author} {\bibfnamefont{W.~J.}\ \bibnamefont{Briscoe}}, \bibinfo {author}
  {\bibfnamefont{M.~W.}\ \bibnamefont{Paris}},\ and\ \bibinfo {author}
  {\bibfnamefont{I.~I.}\ \bibnamefont{Strakovsky}}}%
   (\bibinfo {year} {2011}),\
  \Eprint{http://arxiv.org/abs/1109.0722}{arXiv:1109.0722 [hep-ph]}%
  \bibAnnoteFile{NoStop}{workman-2011}%
\bibitem{khemchandani-2008}%
  \BibitemOpen
  \bibfield{author}{%
  \bibinfo {author} {\bibfnamefont{K.}~\bibnamefont{Khemchandani}}, \bibinfo
  {author} {\bibfnamefont{A.}~\bibnamefont{Martinez~Torres}},\ and\ \bibinfo
  {author} {\bibfnamefont{E.}~\bibnamefont{Oset}},\ }%
  \bibfield{journal}{%
  \Doi{10.1140/epja/i2008-10625-3}{\bibinfo {journal} {Eur.Phys.J.}}\ }%
  \textbf{\bibinfo {volume} {A37}},\ \bibinfo {pages} {233} (\bibinfo {year}
  {2008}),\ \Eprint{http://arxiv.org/abs/0804.4670}{arXiv:0804.4670 [nucl-th]}%
  \bibAnnoteFile{NoStop}{khemchandani-2008}%
\end{thebibliography}%

\end{document}